# The development of a fuzzy regulator with an entry and an output in Fislab

Univ.Asist. Simona Apostol
"Tibiscus" University of Timisoara, Romania

**Rezumat**:Articolul de fata este o continuare a articolului "Fislab the Fuzzy Inference Tool-Box for Scilab" si reprezinta aplicatia practica :"Dezvoltarea unui regulator Fuzzy cu o intrare si o iesire in Fislab". Articolul mai contine pe langa aplicatie si cateva functii care vor fi utilizate in program si anume functii Scilab pentru fuzzificarea informatiei ferme, functii pentru operatia de defuzzificare cat si functii pentru implementarea schemei de inferenta.

## 1. Scilab functions utilize for firm information fuzzyfication

In **fuzzyfication** process of firm information (the input information in RG-F) is necessary for run the next levels:
- definition of base set which characterize linguistic variable (VL);the base set defines variation domain for firm information;
- definition for linguistic terms(TL)which wish to characterized evasive firm information;

this thing supposes apartenence function definition which characterize TL;
- determination of apartenention grads for firm values at TL result definition n apartenetion grads.

   Elementary Fislab functions which are utilized in this intent are:

**Function fzfir**
Function **fzfir** of Fislab, correspond at Matlab function: **addmf.**
**File:** *fzfir.sci*
**Description:** This function implements the operation of fuzzyfication with the effect of defining five types of apartenence functions for the linguistic terms (TL) related to a certain VL.

115

The function admits as entry parameters the followings:
- $A_0$ – the firm value of VL;
- U – the discourse's universe introduced as a vector.

This function returns a vector with belonging's grades of the TL's apartenence function in discussion.

The function's syntax is the following:
ap = fzfir(n,U,a,b,c)

**The mfplot function:**
*Remark:* **Mfplot function** from Fislab is not the same with mfplot function from Matlab, even if they both have the same name.
**File:** *mfplot.sci*
**Description:** This is a function which posts up the linguistic terms(belonging's functions),which are activated in the case of an inference fuzzy matrix or shows only TL's specific to a VL and defined on the U discourse's universe. The function can also graphically represent fuzzy's ordering from a RG-F rule's basics with maximum four inputs and one output.

The function's syntax is the following:
mfplot(a,b,c,d,e,f,g,h,i,j)

Parameters significance is detailed presented in [Ort97].

## 2. Scilab function used for the implementation of inference scheme

It should be mentioned that this is relating to Rg-F Mamdani and for the inference scheme it is implicitly accepted the inference (method) Max-Min by Mamdani (for examples, [PP97]).

**Cri Function**
**File:** *cri.sci*
**Description:** This function implements the inference scheme MAX-MIN of Mamdani, based on a fuzzy inference matrix R, corresponding to the rule base of fuzzy ordering and a *Ap* vector, containing belonging grades of the apartenence function appropriate to the T.L of entry VL. The function returns a *Bp* vector, containing belonging grades of apartenence function relating to TL of output VL.

The function syntax is the following:
[Bp] = CRI(R,Ap)

An example of applying this function is presented as it follows :
```
-->B=[0  0 .1 .5 1];
-->A=[1 .5 .1 0  0];
-->R=[0 0 .1 .5 1;0 0 .1 .5 .5;0 0 .1 .1 .1;0 0 0 0 0;0 0 0 0 0]
 R  =
!  0.   0.    .1    .5   1. !
!  0.   0.    .1    .5    .5 !
!  0.   0.    .1    .1    .1 !
!  0.   0.   0.    0.    0. !
!  0.   0.   0.    0.    0. !
-->[Bp] = cri(R,Ap)
 Bp  =
!  0.   0.    .1    .1   1. !
```

## 3. The function used for defuzzyfication

**Defzfirg function**
**File:** *test.sci*
Defuzzyfication function it is presented as it follows:
*function [y]=defzfirg(x,ap)*
*n=length(x);*
*sum1=0;*
*sum2=0;*
*for i=1:n*
*sum1=sum1+ap(i);*
*sum2=sum2+x(i)\*ap(i);*
*end*
*y=sum2/sum1;*
*endfunction*

For the defuzzyfication procedure from Scilab, there was developed the following syntax of defzfirg:
    y = defzfirg(x,fdap),
in which:
    x – the column vector of independent variable, on which it is defined the linguistic variable (of output);
    y – the firm (scalar) value obtained by the means of;

117

fdap –the column vector of the apartenence function values related to the linguistic variable which was defuzzyficated (command/output).

In the domain of fuzzy there are various defuzzyfication procedures. For this paper it was used the defuzzyfication type of „centre of weight" .The centre of weight method has advantages (all the activated rules work on the realization of the firm command $u_0$ moderated with the specific realization grade), but it also has disadvantages (calculating the weight centre might be a large time consummator or might need realization of an expensive hardware) ([PP97]).

## 4. The development of a fuzzy regulator with an entry and an output in Fislab

It will be next described the implementation in Fislab of a RG-F Mamdani Type for temperature adjustment. In order to achieve this ,RG-F admits an entry VL, "Temperature", T. So the module of fuzzyfication for Rg-F temperature admits as an entry the signal "Temperature" to whom it is associated VL "temperature" and for whom five TL are defined; named{TFJ,TJ,TM,TI,Ffi} and they have apartenece function illustrated in the fig 1.

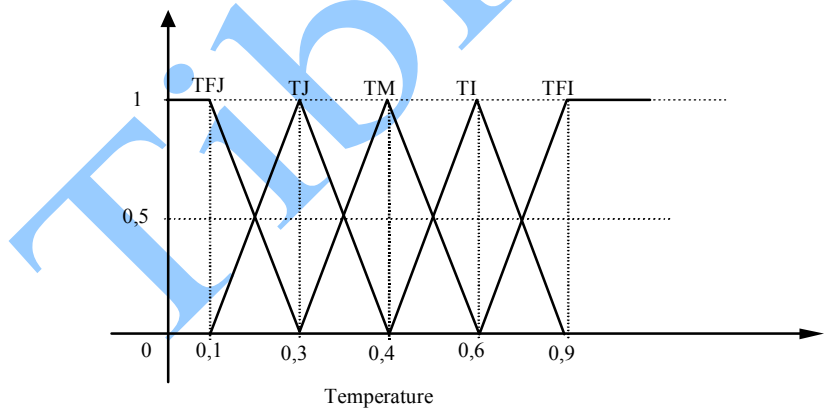

*Fig. 1. Apartenence function related to the entry TL &VL*

It is considered the output „command" u, with the 5 TL named {CVS, CS, CM, CB, CVB} having various apartenence function illustrated in fig. 2.

118

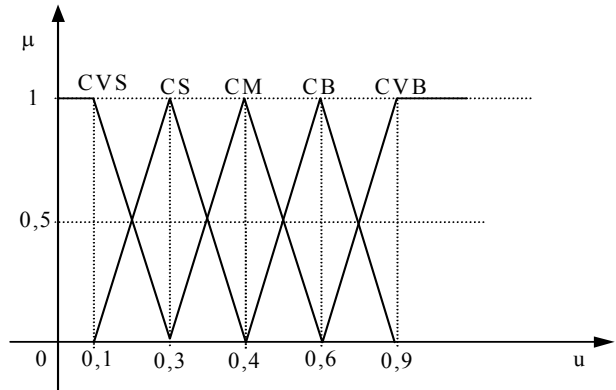

***Fig. 2. Apartenence function of input TL and VL***

For inference it is used the method MAX-MIN of Mamdani sustained by decision table from table 4.3(the rule base is formed by 5 rules).

***Table.1. The rule base of RG-F with an input and an output***

| T | TFJ | TJ | TM | TI | TFI |
|---|-----|----|----|----|-----|
| u | CVB | CB | CM | CS | CVS |

The source code of the program related to this example is illustrated in:

*$A_0=0.7$;*
*$n=3$;*
*$l1=fzfir(n,s,t,A_0)$;*
*$l2=fzfir(n,s,t,A_0)$;*
*$l3=fzfir(n,s,t,A_0)$;*
*$l4=fzfir(n,s,t,A_0)$;*
*$l5=fzfir(n,s,t,A_0)$;*
*[A,B]=rulebase(1,U,a,b,c,1,V,ahat,bhat,chat);*
*g1=cri(R1,l1);*
*l2=A(2,:);*
*g2=cri(R2,l2);*
*l3=A(3,:);*
*g3=cri(R3,l3);*
*l4=A(4,:);*

```
g4=cri(R4,l4);
l5=A(5,:);
g5=cri(R5,l5);
x1=union(g1,g2);
x2=union(g3,g4);
x3=union(x1,x2);
x4=union(x3,g5);
y=defzfirg(x4)
```

In the case of a firm value of input (temperature, noted by $A_0$) the command specific to this can be determined through the program by gradually following the steps related to the modules from RG-F structure:
- the firm value $A_0$ is passed through the fuzzyfication module (using fzfir function);
- the result obtained as a result of fuzzyfication is applied to the inference scheme sustained by the rule base (the rule base function must be used);
- the result obtained as a fuzzy command is then processed by the defuzzyfication module in order to get the firm command (by using defzfirg function)

The program had been tested for various firm values of entry.
The obtained values of the firm commands meet the case of accepted fuzzy rule base and as a result, it validates the correctness of program.